# Diverse patterns of pebbles on sand on Mars and Earth

Zheng Zhu[1], Bernard Hallet[2], András A. Sipos[3,4], Gábor Domokos[3,4], Aileen Yingst[5], and Quan-Xing Liu[6*]

**AFFILIATIONS**

1. School of Ecological and Environmental Sciences, East China Normal University, 200241 Shanghai, China.
2. Department of Earth and Space Sciences and Quaternary Research Center, University of Washington, Seattle, WA 98195, USA.
3. Department of Morphology and Geometric Modeling, Budapest University of Technology and Economics, Műegyetem rkp 3, H-1111 Budapest, Hungary.
4. HUN-REN-BME Morphodynamics Research Group, Budapest University of Technology and Economics, Műegyetem rkp 3, H-1111 Budapest, Hungary.
5. Planetary Science Institute, Tucson, AZ 85719, USA.
6. School of Mathematical Sciences, Shanghai Jiao Tong University, Shanghai 200240, China.

***CORRESPONDING AUTHORS**. Email: qx.liu@sjtu.edu.cn ;
Tel Phone: 086 13661642723

## Key Points

Main point #1:
We revisit one of the planet's most enigmatic early discoveries: the uniform dispersion of clasts, too large to be entrained by wind.

Main point #2:
We illustrate the uniform clast dispersions with high resolution rover images, and analyze them in unprecedented detail and rigor.

Main point #3:
We present numerical model results that reveal 4 distinct patterns and elucidate underlying processes, helping resolve the enigma.

**Abstract (197 words)**

On Mars, fields of sand dunes contrast with the general cratered, rocky terrain commonly seen from orbit. Near the equator, in Gale Crater, images from the rover, Curiosity, also reveal order on smaller scales: ripples on dunes, and ground patterns in scattered sites. The patterns include relatively inconspicuous forms: evenly spaced pebble-size rocks (termed clasts) on meter-scale domains of wind-blown sand. Here, we examine quantitatively several such domains on both Mars and Earth. The domains are significantly more orderly than expected by chance. Moreover, many are "hyperuniform", a self-organized state recently recognized in diverse active materials and biological systems but that appears novel for planetary surfaces. We use numerical simulations to examine how diverse clast distributions, ranging from random and hyperuniform dispersions to distinct alignments, can emerge spontaneously from clast displacements induced by gravity, combined with the wind-driven evolution of the surface, sand transport, and ripple migration. This paper highlights easily overlooked self-organized patterns beyond distinct geometric patterns on at least two planets, and the simulations help understand the information coded in clast domains. Moreover, our methods and findings potentially have quantitatively implications for studies of issues of global significance on Earth, including dust emission from vast areas into the atmosphere.

**Plain Language Summary**

On both Earth and Mars, the surfaces of sandy landscapes often feature patches of scattered pebbles. While these may appear randomly placed, in some regions the pebbles are arranged in surprisingly uniform and organized patterns. Using high-resolution images from NASA's Curiosity rover on Mars and drone photos from the Gobi Desert in China, we analyzed how these pebble patterns form. Our study shows that the pebbles can self-organize into a rare state called "disordered hyperuniformity", which has only recently been recognized in physical and biological systems. To understand how these patterns emerge, we developed a numerical model that simulates how wind-blown sand and moving ripples can shift pebbles into organized patterns over time. These findings suggest that even subtle surface features on Mars carry important information about wind activity and surface processes. They hint that organized pebble patterns might help reduce dust emissions in desert regions, which has implications for understanding climate and environmental changes on both Mars and Earth, and perhaps other planets.

## 1. Introduction

The natural world around us exhibits distinct patterns at all scales in diverse systems, from the intricate arrangement of cells and tissues [*Camazine*, 2001; *Cheng and Ferrell Jr.*, 2019; *Nedelec et al.*, 1997] to the macroscopic organization of ecosystems and landscapes [*Kessler and Werner*, 2003; *Lämmel et al.*, 2018; *Li et al.*, 2021; *Rietkerk et al.*, 2004]. Since the seminal work of Alan Turing [1952] on pattern formation in biological systems a large body of experimental evidence and theoretical work has shown that such self-organized structures generally arise from scale-dependent feedbacks and local interactions, allowing particles -- animate or not-- to behave in surprisingly coordinated ways [*Camazine*, 2001; *Kondo and Miura*, 2010; *Meinhardt*, 2009; *van de Koppel et al.*, 2008]. Theoretical models, aided by measurements, have shed light on the underlying processes and enabled diagnostic comparisons between observed and simulated patterns [*Kessler and Werner*, 2003; *Lasser et al.*, 2023; *Morison et al.*, 2021; *Rapin et al.*, 2023; *Rodriguez-Iturbe and Rinaldo*, 1997; *Werner and Hallet*, 1993; *Zhao et al.*, 2021]. Yet, dispersions of pebble-size clasts on sand – such as those illustrated in Fig. 1 -- have received little attention compared to geometric patterns that have been extensively studied in ecosystems [*Dong et al.*, 2023; *Klausmeier*, 1999; *Liu et al.*, 2013; *Meinhardt*, 2009; *Murray*, 2003; *Zelnik et al.*, 2015] and geomorphic systems on both Earth [*Bagnold*, 1941; *Corte*, 1963; *Goldthwait*, 1976; *Hallet*, 2013; *Kessler and Werner*, 2003; *Lämmel et al.*, 2018; *Li et al.*, 2021; *Rodriguez-Iturbe and Rinaldo*, 1997; *Tholen et al.*, 2022; *Washburn*, 1956; *Werner*, 1995; *Werner and Hallet*, 1993] and Mars [*Bridges and Ehlmann*, 2018; *Morison et al.*, 2021; *Rapin et al.*, 2023; *Telfer et al.*, 2018; *Trowbridge et al.*, 2016]. Prior research has investigated distinct geometric patterns formed of larger rock fragments on surfaces of loose rock debris [*Barrett et al.*, 2017; *Hallet et al.*, 2022; *Ward et al.*, 2005], but little is known about the origin and information coded in the subtle organization of pebble-size clasts scattered on sand on Mars. These fine-scale patterns invite a deeper look at this organization and underlying processes that generate and shape them.

Distinct self-organized forms -- sand dunes and ripples, and geometric patterns of clasts (stripes, circles, and labyrinths) – have been well-recognized on planetary surfaces [*Bridges and Ehlmann*, 2018; *Hallet et al.*, 2022; *Malin et al.*, 1998; *McGill and Hills*, 1992; *Rapin et al.*, 2023], including Mars and Earth. At the Spirit landing site on Mars, the uniform dispersion of clasts, too large to be entrained by wind, was one of the planet's most enigmatic early discoveries [*Pelletier et al.*, 2009; *Ward et al.*, 2005]. More uniformly spaced distributions were found in all areas than expected under Poisson distribution models, with one exception: especially coarse and recent impact-ejecta deposits, presumably because insufficient time had elapsed since emplacement for the surface ejecta to evolve and self-organize [*Pelletier et al.*, 2009]. These data were utilized, along with results from wind-tunnel experiments and numerical simulations, in the pioneering, comprehensive study of clast distributions on Mars [*Pelletier et al.*, 2009]. On Earth, similar evenly spaced clasts on sand surfaces have been observed and puzzled over for decades, extending back at least to Bagnold's [*Bagnold*, 1941] early descriptions of the tendency of pebbles on sand to disperse "*to the most uniform distribution possible*". Such distributions are not limited to modern aeolian surfaces; for instance, Walker and Harms [*Walker and Harms*, 1972] illustrated examples from the ancient stratigraphic record (e.g., their figure 16).

Much insight into clast distributions can be gained from recent advances in studies of active matter -- systems composed of energy-driven particles that interact dynamically with their surroundings -- across various fields that recognize a significant type of spatial distribution, known as hyperuniform [*Donev et al.*, 2005; *Hexner et al.*, 2017; *Torquato*, 2018; *Torquato and Stillinger*, 2003]. In mixed-component, active-particle fluid systems (here, "active" refers to particles that consume energy from their surroundings to generate motion) [*Backofen et al.*, 2024; *B Zhang and Snezhko*, 2022], hyperuniform distributions are characterized by unusually low fluctuations in component density at large scales, especially when compared to disordered systems such as mixed ideal gases or typical liquids [*Lei et al.*, 2019]. Some of these, those referred to as disordered hyperuniform systems, are statistically isotropic; they result in distributions that, at first glance, may appear random, but behave in a relatively organized manner over large distances [*Torquato*, 2018; *Torquato and Stillinger*, 2003]. Most commonly hyperuniform systems are characterized by a structure factor $S(\boldsymbol{k}) \sim |\boldsymbol{k}|^{\gamma}$ with $\gamma > 0$ for $|\boldsymbol{k}| \to 0$ ($\boldsymbol{k}$ being the wavevector and $|\boldsymbol{k}|$ the wavenumber), with different classifications depending on scaling exponent, $\gamma$ [*Torquato*, 2018]. Here, the term wavenumber simply means a spatial frequency that describes periodic variation in space; it is the inverse of a characteristic length scale of the spatial pattern. Hyperuniform states have gained considerable attention across physical [*Chremos and Douglas*, 2018; *Hexner and Levine*, 2015; *Lei et al.*, 2019; *Torquato et al.*, 2015; *Xie et al.*, 2013] and biological systems [*Huang et al.*, 2021; *Jiao et al.*, 2014; *Lubensky et al.*, 2011; *Torquato*, 2021], largely due to their common optimal properties. For example, they can suppress diffraction in photonic materials [*Man et al.*, 2013] and minimize free energy in two-phase heterogeneous materials [*Torquato*, 2016]. Marine algal cells provide a well-studied example in the biological realm; they can self-organize into disordered hyperuniform distributions at air-liquid interfaces, evidently optimizing nutrient uptake through repulsive interactions [*Huang et al.*, 2021; *Zhu and Liu*, 2019]. Despite the profound connections between hyperuniform states and numerous areas of fundamental science, their implications for the wide range of patterns in natural landscapes remain largely unexplored (e.g. Fig. 1).

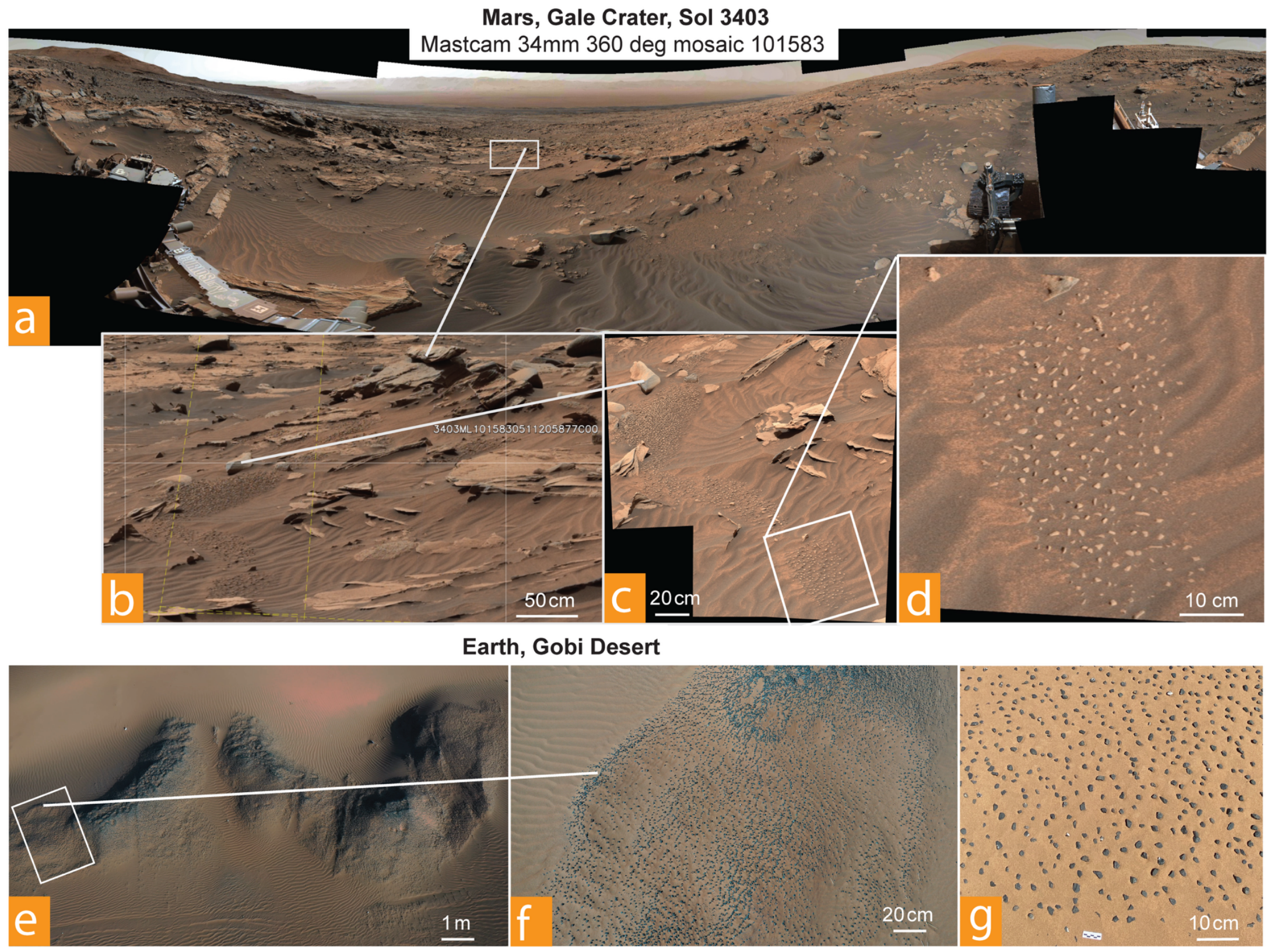

**Fig. 1. Context for the patches of uniformly distributed pebble-sized clasts on aeolian sand on Mars and Earth**. (**a**) Panorama of the Mars landscape in Gale Crater on Sol 3403. (**b**) A close up of the area outlined by the white rectangle in image (a); note the scale bars for the subframes. (c) Enlargement of the region near the rock at ends of the white line in both images (a) and (b). Rectangle outlines location of the (d) subframe. (**d**) A patch of pebbles showing a distinct uniformity in their spacing and size (Sol 3405). (**e**) Drone (DJI Mavic 3T) image showing an overview of a study site in the Gobi Desert, China (see Materials and Methods). The field of view spans approximately 30 m. (**f**) Close up of the white rectangle area in image (e). (**g**) Patch of pebbles on Earth, similar to subframe (d) from Mars, displaying clear uniformity. The scale bar is 10 cm. Image credits: NASA/JPL-Caltech/MSSS (a-d) and (e-g) Quan-Xing Liu.

In this study, we document and analyze in detail the spatial distributions of clasts on sandy surfaces interspersed in rocky terrain using MastCam images acquired from the Mars Science Laboratory (MSL) rover, Curiosity [*Malin et al.*, 2017] in Gale Crater, on the red planet. For comparison, we also examine clasts from the Gobi Desert on Earth. Here, "Gobi" means a desert that differs from "sand desert", presumably a stone (or stony) desert. At several sites on both Mars and Earth, surface clasts appear uniformly dispersed over meter-scale sand domains (Fig. 1 and Figs. S1—S4). These domains are widespread along Curiosity's ~ 36 km-long traverse to date (September, 2025) and they have been reported from other Martian rover sites (e.g. Fig. 2d in Ref. [*Yingst et al.*, 2008]), but they are relatively rare. Importantly, these contrast with surfaces commonly imaged on Mars that consist of 1) regolith (scattered, loose rock fragments of various sizes and spacings) and bedrock exposures, and 2) dunes and smaller aeolian bedforms that generally occur as extensive fields of well sorted sand with coarser grains commonly highlighting bedform crests [*Sullivan et al.*, 2022].

Our work on clast distributions benefits from, and complements the extensive literature about clasts at Gale Crater that documents and seeks to understand the size, geometry, composition and geologic history of sands and clasts [*Ehlmann et al.*, 2017], and how these surfaces of loose material would appear from orbit. For instance, Yingst et al. [*Yingst et al.*, 2008] analyzed morphologic characteristics of surface rock particles at five different landing sites (Viking 1 and 2, Mars Pathfinder, Spirit at Columbia Hills, and Curiosity at Gale Crater) to classify clasts morphologically and to potentially infer the origins and transport history of discrete clast populations. This broad interest in clasts on Mars has led to considerable geologic context and extensive documentation, including an ongoing program of systematic imaging (Clast Surveys) in Gale Crater [*Yingst et al.*, 2013], which continues to benefit the type of work presented herein.

We leverage the seminal studies of clast distributions [*Bagnold*, 1941; *Pelletier et al.*, 2009; *Ward et al.*, 2005], and augment them in several ways: 1) using an extensive set of recent, higher resolution Mars images from the Curiosity rover to document in detail clast distributions on sandy surfaces within Gale Crater, and using Earth images from the Gobi Desert for comparison; 2) characterizing the clast distributions rigorously with modern quantitative methods; 3) interpreting them from the rich, contemporary perspective of hyperuniformity; 4) presenting a new model of aeolian redistribution of clasts on sand that includes the significant role of migrating ripples and yields four distinct distributions that closely resemble those in nature, and 5) introducing a phase diagram as a visual synthesis of the modeled distributions.

## 2. Materials, Methods, and Metrics

### 2.1 Data set

The primary Mars data utilized in this work are images acquired on the Martian surface by the Mars Science Laboratory (MSL) rover's Mastcam stereo imaging system [*Malin et al.*, 2017]. This system consists of two cameras with different focal lengths (Left, 34 mm, and Right, 100 mm) mounted side-by-side on the rover's Remote-Sensing Mast, about 2 m above the ground surface. Malin et al. [2017] provide detailed information on the camera, as well as on the associated hardware and software used to produce the color images. This work also utilizes image data from the arm-mounted Mars Hand Lens Imager MAHLI, [*Edgett et al.*, 2012], which provides close-range, hand lens-quality images when placed near ground targets.

Labels for the MSL images and mosaics generally start with the date of image acquisition as the sol number (Martian day of the MSL mission, starting Aug 6, 2012), followed by the camera's focal length (ML or MR) and the observation's sequence number. Figure 1, for example, features a 360-degree panorama of ML images "sol03403_ML_101583" acquired on sol 3403 using the 34 mm focal length camera. The Mastcam panorama "ML_101583" comprises 126 individual images.

Comparable images of analog Earth sites were taken with a hand-held digital camera and, for the overviews, a Drone camera with focal length 24 mm and aperture f/2.0, and field of view 84°. The methods used are summarized in the next section.

### 2.2 Data collection

The MSL rover images used in this work are archived for free public access by the NASA Planetary Data System (PDS). Within the PDS, MSL image data are available at https://pds-imaging.jpl.nasa.gov/volumes/msl.html (e.g., Mastcam data at https://doi.org/10.17189/1520328). The clasts images from Mars can also be downloaded from https://mars.nasa.gov/msl/multimedia/raw-images/. Image numbers and details are listed in Tables S1 and S2. The mosaics used in Figs. 1 and 5e were produced by Malin Space Science Systems (MSSS) using MSL image data (images acquired by NASA/JPL-Caltech/MSSS). Note that although many MSL images include clasts along the Curiosity trajectory [*Hallet et al.*, 2022], many do not display distinct uniform clast domains, resulting in a limited total number of relevant images. Only some of the images used here are included for precise analysis.

The images of clast patches on Earth used in this work are from study sites in the Gobi Desert in northwestern China, 38°42′6″ N, 105°14′18″ E (Tengger, Alxa Left Banner, Figs. 1e to G and Fig. S3). These images were taken in the western region of the Gobi Desert, adjacent to the Tengger desert, where dispersed clasts on fine sand are common. We present one of the largest clast patches observed during our field investigation, using high-resolution images (Fig. S3a and Fig. 1e) taken with a DJI Mavic 3T drone 20 meters above the ground with a resolution of about 0.4 cm·per pixel. The mean slope of the area with ripples in this figure is ~9° based on high precision RTK (Real-Time Kinematic) elevations.

### 2.3 Image analysis

We first convert the color scenes to grayscale images, and extract the clast center points after binarizing, using MATLAB and Otsu's method [*Otsu*, 1979], which chooses the threshold value to minimize the intraclass variance of the threshold black and white pixels. We, then: 1) Use MATLAB to correct the positions where the density of clasts is high, because two close clasts might be processed as one due to shadows. Although we have adjusted the threshold for the binarization, several clasts still require custom-code correction. 2) Determine density fluctuations from the data with different window sizes (e.g.: sol 2312: $l = 12$ to 45 mm; sol 3303: $l = 8$ to 30 mm; sol 3403: $l = 4$ to 10 mm). All original MATLAB programs used for position tracking have been adapted from existing literature focused on dense mussel beds [*van de Koppel et al.*, 2008].

### 2.4 Comparison of the data with distributions generated with the Poisson point process

Let $f_o^{emp}(\cdot)$ and $f_s^{emp}(\cdot)$ denote, respectively, the empirical probability density associated with the observed and simulated data (Poisson process) weighted by the sample mean. After removing the edges and triangles adjacent to the domain boundary, we computed the empirical distributions of the triangle *areas* (i.e., $f_o^{emp}(A)$ and $f_s^{emp}(A)$) and edge *lengths* (i.e., $f_o^{emp}(L)$ and $f_s^{emp}(L)$). Our null hypothesis is that the two samples are drawn from a continuous distribution. The two-sided Kolmogorov-Smirnov tests at the $\alpha = 5\%$ significance level yield $p = 1.02 \times 10^{-19}$ and $8.23 \times 10^{-8}$ for the asymptotic probabilities for the area and the edge length distributions, respectively. As $p < 0.05$, we reject the null hypothesis for both the area and the edge length distributions and conclude that there is strong evidence that the distributions generated by Poisson process and the observed data differ from one another significantly.

To compare observed and simulated distributions further (Fig. 2, a and b), we use the probability density function to obtain a smooth estimate of the discrete data on nearest neighbor distance, edge length, and area (Fig. 2, c to e). The area distributions $f_o^{emp}(A)$ and $f_s^{emp}(A)$ both follow Gamma distributions, whereas the edge lengths $f_o^{emp}(L)$ and $f_s^{emp}(L)$ are more consistent with Weibull distributions (thick solid lines). The best-fit parameters (denoted as *a* and *b*) of these two-parameter distributions [*Forbes et al.*, 2011] and their 95% level-confidence intervals are summarized in Table S3.

**Table 1.** Spacing measures of clast distributions in Gale Crater, Mars: Std refers to standard deviations, and PPP to distributions produced by the Poisson point process. Data were normalized by their mean values. $R$ is the spacing measure defined by Ward et al [*Ward et al.*, 2005]. An $R < 1$ indicates a clustered distribution ($R = 0$ would mean that all clasts are on top of one another), $R = 1$ corresponds to a random distribution, and $R > 1$ indicates a dispersed distribution tending toward uniformity. For reference, $R = 2.0$ for a square lattice, and the theoretical maximum dispersion is $R = 2.149$ for a triangular lattice [*Barrett et al.*, 2017].

| | Sol 3405 | Sol 2312 | Sol 3403 | Sol 3409 | PPP |
|---|---|---|---|---|---|
| $R$ | 1.35 | 1.28 | 1.27 | 1.46 | 1.00 |
| Std (min_distance) | 0.34 | 0.34 | 0.31 | 0.26 | 0.53 |
| Std (edge_length) | 0.40 | 0.44 | 0.61 | 0.41 | 0.49 |
| Std (area) | 0.60 | 0.67 | 1.58 | 0.59 | 0.86 |

We note that the pairwise regions *spanned* by the confidence intervals of the parameters $a$ and $b$ are nonoverlapping for both the $f_o^{emp}(A)$, $f_s^{emp}(A)$ and the $f_o^{emp}(L)$ and $f_s^{emp}(L)$ pairs. This demonstrates geometrically that the recorded data indeed differ from a dataset generated by Poisson point processes. We compare the triangulated meshes associated with both datasets and find that the variances both of the triangle areas and the triangle edge lengths are smaller in the investigated empirical dataset (Table S4 and Fig. 2). As we found the same result for all studied images, we conclude that the uniformity in the studied images of clasts all surpass the uniformity of the results from Poisson point processes (Tables 1 and S5, and Figs. S4 and S5), which is a strong indicator that the clast distributions are organized, rather than produced by chance.

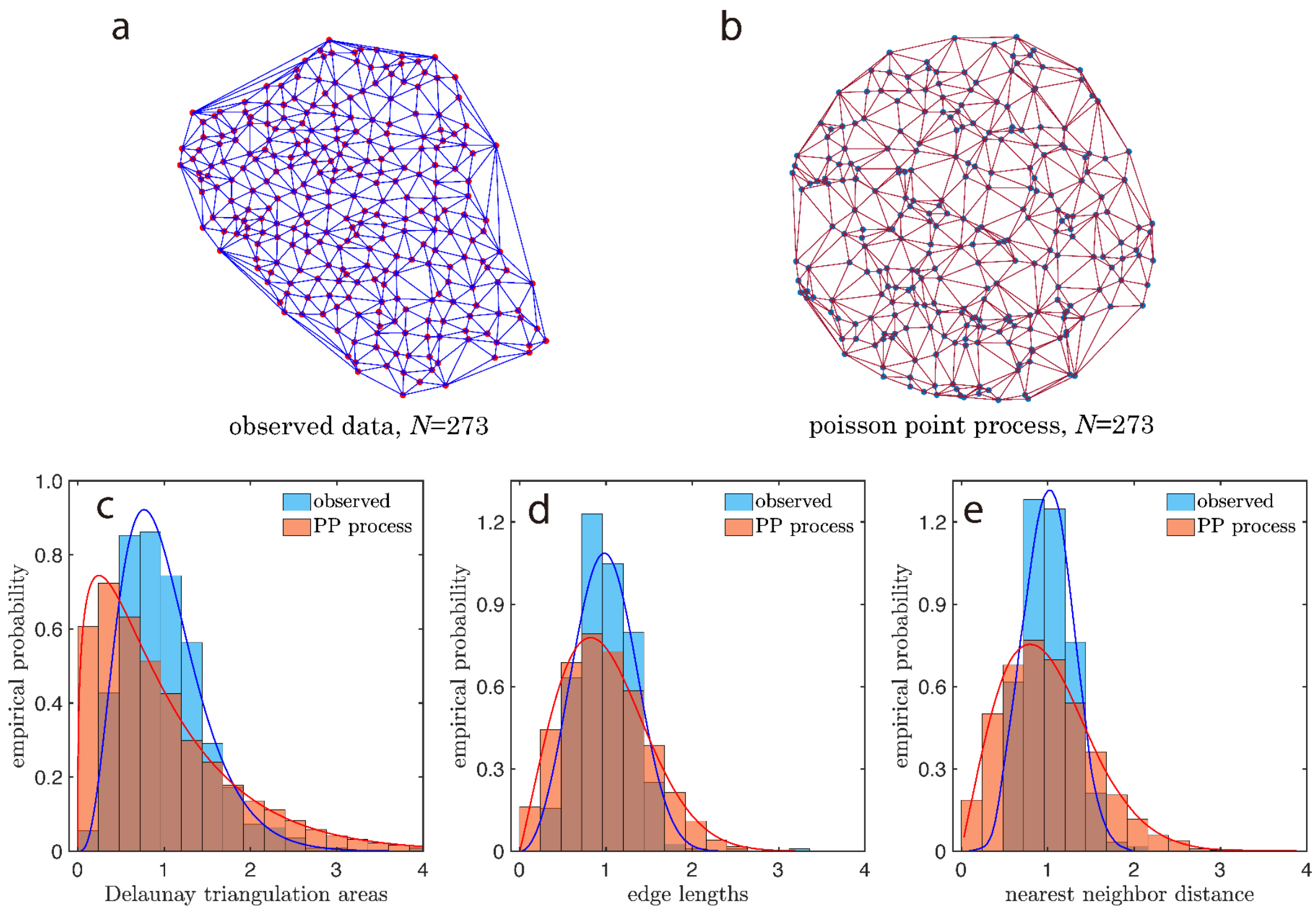

**Fig. 2. Comparison of observed clasts (blue in all insets) to random points generated from the Poisson process (orange).** (**a**) Delaunay triangulations for the 273 clasts imaged on sols 3403 and 3405, and as shown in Fig. 1d. (**b**) Delaunay triangulations for random points generated with $N = 273$ vertices. (**c** to **e**) Normalized empirical distributions of the nearest neighbor distances (c), and edge lengths (d), and triangle area (e). The normalization is conducted using probability density function for a Gaussian distribution. Thin solid lines show the best-fit probability density functions (Weibull distribution for the nearest neighbor distance and edge length; Gamma distribution for the triangle area). Similar results were obtained for other sites (see Figs. S4 and S5).

### 2.5 Cellular model of clast displacements on a dynamic sand surface.

The model consists of two parts: the development of the surface of wind-erodible sand and clast displacements caused by the wind and gravity over that surface. To simulate the sand surface simply, we adapt a cellular model of eolian dune development similar to Werner [1995]. Its fundamental elements are: 1) a square lattice with an initially uniform sand thickness (such as 15 sand grains) on a horizontal surface; 2) randomly chosen units (a volume much larger than the grain size) of sand on the lattice are transported downwind a distance $l$; and 3) at the new location, the sand unit is deposited with a probability $p$ ( $p < 1$ and if it is in a shadow zone protected from the direct impact of saltating grains, where $p = 1$). The values of $p$ and $l$ effectively represent the local wind speed that controls the distance of sand transport and rates of deposition and erosion, higher values of $l$ (longer saltation distance) and lower values of $p$ (smaller deposition probability) lead to faster transport. Shadow zones are defined as areas not directly impacted by saltating grains; they would not receive direct light if the surface were illuminated by an upwind light source angled downward 15° from the horizontal. The down-wind length of shadow area is about $1/\tan 15° \approx 4.0$. Moreover, if a sand unit is not deposited where it lands, it continues downwind a distance $l$ until deposition occurs. Lastly, because of net deposition, the sand surface rises at the new location, and conversely it drops where there is a net divergence of sand. Due to the inferred lack of cohesion of the sand, the maximum surface slope is the angle of repose (i.e., assumed to be ~30°). In this simulation, periodic boundary conditions are used in two-dimensions.

To study closely the evolution of clast distributions, we coupled the simulated dynamics of the ripples to the displacement of clasts on the surface much as was done in the published model for smooth sand surfaces without ripples [*Pelletier et al.*, 2009]. Clasts resting on sand enhanced erosion on their windward side, so in these simulations, the deposition probability of the foreshadow zone of the clasts is zero and the shadow zones comprise two parts, clasts and sand. In areas that belong to both shadow zones and foreshadow zones, according to the net depositional effect of obstacles [*King et al.*, 2005], the shadow zone takes precedence; it is represented by $p = 1$. Clasts roll down slope when the local bed slope beneath the center of each clast exceeds a threshold value given by $\Psi$, $\frac{\partial h}{\partial x,y} \geq \tan \Psi$. They continue rolling downhill until the local slope decreases below this threshold. According to a previous study [*Miller and Byrne*, 1966], the angle $\Psi$ of repose or stability for a single clast is $\Psi = \alpha \left(\frac{D}{K}\right)^{-\beta}$, $D$ is the clast diameter, $K$ is the mean diameter of the substrate (sand grains). Assuming that $\beta \sim 0.3$, $\alpha \sim 40°$ and $D$ is ~7 times larger than sand grains, $\Psi$ is about $40° \times 7^{-0.3} \sim 22°$, which corresponding to a max sand height difference about $\tan 22° \times \frac{7}{2}(1 + \sin 22°) \sim 2.0$ units.

In short, in the simulations the probability $p$ of deposition exerts the primary control on the formation of ripples. The local surface steepening due to ripple migration and erosion on leading side of pebbles destabilizes them, causing some to tumble.

### 2.6 Metrics

**Pair correlation function.**

Pair correlation functions $g_2(r)$ define the average number of particles surrounding a reference particle. It is a simple summary statistic that can be used to characterize a tendency toward aggregation (small distances $r$ that $g_2(r) > 1$) or dispersion [*Wiegand and Moloney*, 2013]. It is calculated as $g_2(r) = \frac{1}{N\rho} \langle \sum_i \sum_j \delta(\boldsymbol{r}_{\mathrm{ji}} - r) \rangle$ and $(i \neq j)$, where $\rho$ is the number density of particles, $N$ is the total number of particles, $\langle\ \rangle$ denotes the ensemble average, $r_{ji}$ is the distance between particle $i$ and particle $j$. The position of the first peak denotes the average distance between particles; its height (magnitude) represents the tendency of particles to be found at that distance from each other more frequently than at elsewhere.

**Density fluctuation and hyperuniform distributions.**

Density fluctuations describe how density variations at different points in space are related to each other. In two dimensions, density fluctuations are calculated as $\langle \Delta\rho^2 \rangle \equiv \langle \rho^2(\ell) \rangle - \langle \rho \rangle^2$ for square interrogation windows of different sizes $\ell$, $\langle\ \rangle$ denotes the ensemble average and $\rho$ is the average density of the system. Note that $\langle \Delta\rho^2 \rangle \sim \ell^{-2}$ represents a Poisson process. Importantly for this paper, systems whose density fluctuations decay faster than a Poisson process, $\langle \Delta\rho^2 \rangle \sim \ell^{\alpha}$ with $-3 \leq \alpha < -2$, are called hyperuniform. The more uniform a system is, the larger $|\alpha|$ is. Systems resembling perfectly periodic arrays like crystals can achieve or closely approach the maximum absolute value: $|\alpha| = 3$ [*Torquato*, 2018].

**Structure factors.**

Structure factors, $S(k)$, are useful to describe the distribution of particle density in systems over a range of spatial scales, represented by the size, $\ell$, of observation windows. They provide information about the spatial correlations (e.g. positions, ordering, and fluctuations) between particles in 2D space. It is calculated as:

$$S(\mathbf{k}) = \frac{1}{N}\langle | \sum_{j=1}^{N} e^{-i\boldsymbol{k}\cdot\boldsymbol{r}^{(j)}} |^2 \rangle,$$

where $N$ is the total number of particles, $r$ is the particle position, $\mathbf{k}$ is the wave vector, $\boldsymbol{k} = 2\pi/\ell$ and $|\boldsymbol{k}|$ for the wavenumber. In two dimensions, systems whose fluctuations decay faster, $S(\boldsymbol{k}) \sim |\boldsymbol{k}|^{\gamma}$, with $0 < \gamma \leq 1$ for $|\boldsymbol{k}| \to 0$, are called hyperuniform [*Torquato*, 2018]. Additionally, perfectly periodic arrays like crystals can achieve or closely approach the maximum value of $\gamma = 1$. Illustrations and comparisons of several different two-dimensional point configurations can be found by Torquato [2018] and Table 1.

## 3. Results

### 3.1 Clast Distributions

To determine whether the uniform clast distributions in nature could arise by chance, we rigorously compare natural clast distributions to sets of points obtained from homogeneous Poisson point process (herein abbreviated: PPP; Fig. 2a and 2b). Two-sided Kolmogorov-Smirnov (KS) tests [*Conover*, 1972; *Dimitrova et al.*, 2020] reveal a significant difference between clasts and PPP sets (KS test, $p < 0.001$, $Z = 0.224$, $Z$ is the maximum difference between the empirical cumulative distribution functions of the two samples.). There is no overlap of statistical parameters using either edges lengths or areas of the Delaunay triangles (Table S4). These results show that it would be extremely improbable (less than one chance in a thousand) for the observed clast distributions to arise by chance. Several sites yielded comparable results with statistically significant differences between observed clasts and PPP sets. Figures S5 and S6 show six additional clast distributions on Mars and Earth and their PPP counterparts. Their standard deviations and $R$ values, the spacing measure defined by Ward et al. [*Barrett et al.*, 2017; *Ward et al.*, 2005], are listed in Table 1, and Table S5. $R$ is the ratio of the average nearest neighbor distance ($r_A$) to the expected average nearest neighbor distance ($r_E$) of a Poisson distribution. Thus, the observed clast arrangements on Mars and Earth are more spatially uniform than would be expected by chance, based on uniform PPP distributions, suggesting strongly that these clasts are organized.

We use three complementary approaches to characterize the distribution of clasts in greater detail. First, we introduce structure factors, $S(k)$ where $k$ represents the wavenumber (see Materials and Methods for details). The clasts show a scaling behavior typical of hyperuniformity: $S(k)$ increases with the wavenumber ($k$ = 2π/mean clast spacing) , as $S(k) \sim k^{\gamma}$ with $\gamma \sim 0.4$ in the small-$k$ regions (red dashed lines. Fig. 3a and 3d). With increasing $k$, it then fluctuates around ~1.0 (black dashed lines), which is typical of PPP in the large -$k$ regions. It is noteworthy that the exponent value of 0.4 represents a novel example of class III hyperuniformity; as far as we are aware, it has not been reported in geomorphic systems. Here, density fluctuations refer to variations in density within a specific area, measured over spatial scales $\ell$ ranging from 1 to 200 mm.

Second, we characterize the disordered hyperuniformity equivalently in terms of the areal density fluctuations, $\langle \Delta\rho^2(\ell) \rangle$ (Figs. 3b, 3e and S7). The density of pebble numbers grows with the radius $\ell$ of the window under consideration; where it grows more slowly than the window's area (e.g., $\langle \Delta\rho^2(\ell) \rangle \sim \ell^{\alpha}$ , $\alpha \in (-3.0, -2.0)$ for disordered hyperuniform and $\alpha \in [-2.0, 0]$ disordered nonhyperuniform [*Torquato*, 2018]), it reveals the areal density fluctuations of the clasts. Over large length scales, $\ell$, density fluctuations follow the scaling law determined by the central limit theorem: large numbers of independent and identically distributed random variables tend to follow a normal distribution as the sample size increases. Fluctuations first decay nonlinearly (exponent, $\alpha \sim -2.4$) with increasing $\ell$ up to a value above which they stabilize, and then approach an exponent of $\alpha = -2.0$ (Fig. 3b and 3e). In two dimensions, $\alpha = -2.0$ for Poisson point patterns, $-3.0 < \alpha < -2.0$ for hyperuniform distributions, and $\alpha \sim -3.0$ for crystals because of the standard surface-area scaling.

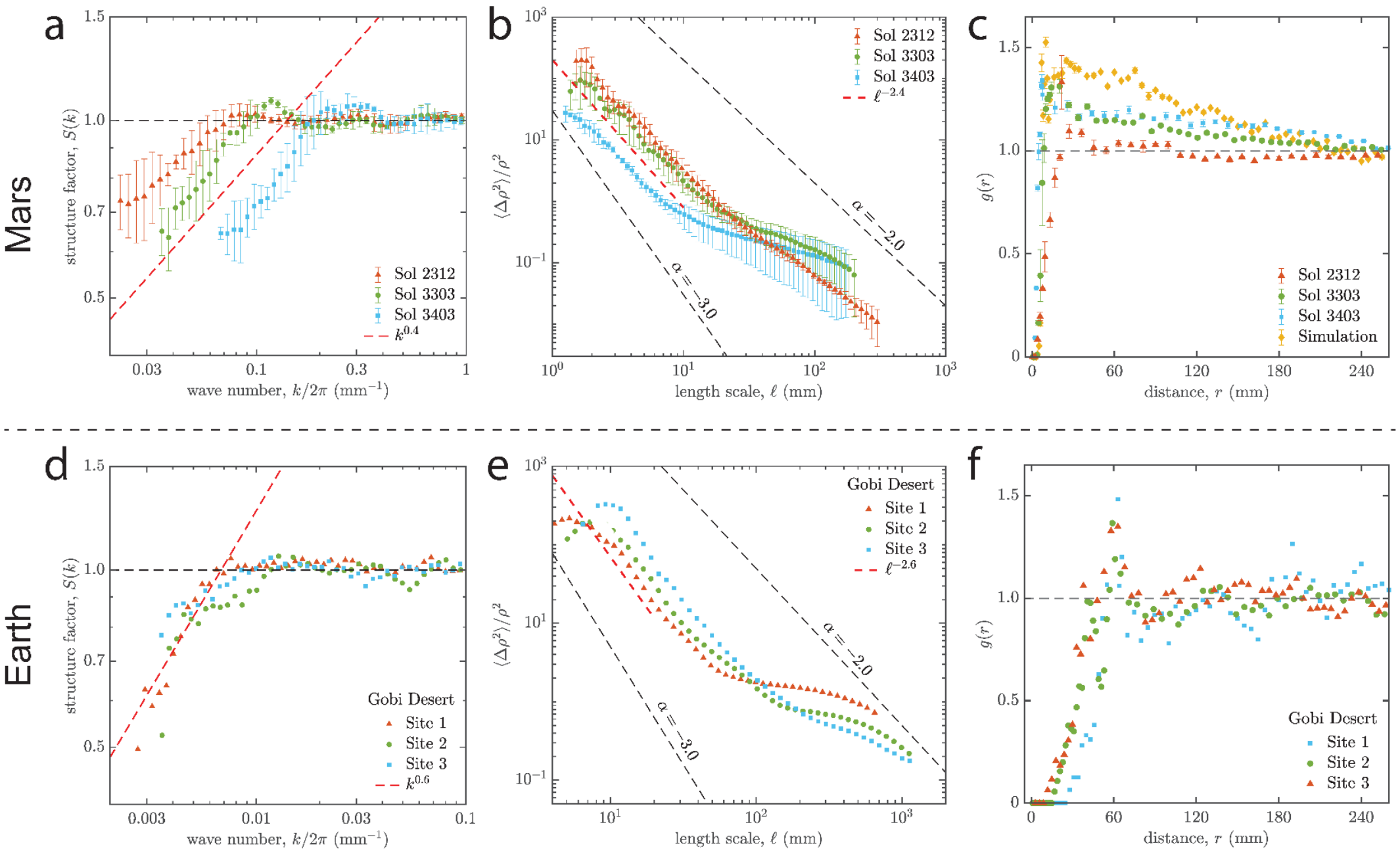

**Fig. 3 Spatial distributions of clasts exhibiting characteristics of disordered hyperuniformity.** (**a**) Structure factors for clasts in the study images in Mars' Gale Crater, shown as a function of wave number for three sites imaged on different dates and locations along Curiosity's route: sol 2312 (orange, adjacent clasts areas, see Table S1), sol 3303 (green), and sol 3403 (blue). Red dashed line represents the scaling law for a disordered hyperuniform distribution with an exponent of $\gamma = 0.4$ for $S(k)$. It delineates that the $k^{0.4}$ corresponds to $\alpha = -2.4$ for density fluctuations. (**b**) Clast density fluctuations

versus study window size, which limits the maximal measurement scale. (**c**) Pair-correlation function $g(r)$ of clasts derived from numerical simulation and observations. Note that all pair correlation functions decay slowly to unity as $r$ increases, implying the existence of long-range correlations between clasts. (**d** to **f**) Corresponding data and results for three Earth sites in the Gobi Desert, reflecting the same analyses as those that produced the Mars panels (a-c). These figures also show the diagnostic characteristics of disordered hyperuniformity, with the conditions $-3 \leq \alpha < -2$. Error bars in panels (a to c) denote one standard deviation (mean $\pm$SD).

Third, we calculate the pair-correlation function $g(r)$ to characterize the spatial arrangement of the clasts more precisely [*Kästner et al.*, 2024]; see the section ***Pair correlation function*** in the supplement for the definition and explanation. In a random distribution, $g(r)$ would equal 1.0 across all scales, indicating the absence of preferred spacings. A distinct peak in $g(r)$ reflects a preferred spacing, defined as the average center-to- center neighbor distance. Such peaks are typical of self-organizing systems, where local order emerges despite the lack of long-range periodicity [*Tarnita et al.*, 2017]. Beyond the peak, $g(r)$ returns toward 1.0, signifying increasing randomness at larger distances. Peaks occur at ~7 mm for the studied clast domains on Mars (Fig. 3c) and 70 mm for those on Earth (Fig. 3f), showing that clasts are more likely to be spaced at these distances than in a random (PPP) distribution, consistent with an organized structure. The clast spacing difference between the Gobi (Tengger, Alxa Left Banner) and Gale Crater sites likely reflects intrinsic differences between sites in wind strength, sand transport dynamics, and/or ground surface properties; these include the fraction of the ground covered by clasts, clast size, and other sediment characteristics. At the Gobi and Mars sites, clasts diameters are generally around 20 mm and 3 mm, respectively (Fig. S2 and S3); correspondingly, their preferred spacings are 70 mm and 7 mm.

For vanishing $k$, the deviation of the structure factor $S(k)$ from hyperuniformity as $k$ decreases to zero, suggests that the clasts do not self-organize into full hyperuniformity across the entire system. However, local hyperuniform regions can extend up to ~10 clast diameters. For instance, the hyperuniform scale can reach 30 mm ($k \sim 0.03$ mm$^{-1}$) for sol 3303 (green line in Fig. 3a), with mean clasts diameters of 3 mm. Similar results were observed on Earth in the Gobi Desert (Fig. 3, d to f). The structure factors and density fluctuations consistently show the hallmark scaling of disordered hyperuniformity (Fig. S8) [*Hexner and Levine*, 2015; *Huang et al.*, 2021; *Torquato*, 2018; *Torquato and Stillinger*, 2003]. We note that the exponent in the small $k$ region reflects the weakest form of hyperuniformity reported in the literature— class III ([*Torquato*, 2018], section 5). For clasts on sandy surfaces, their spatial patterns can arise from periodic forces created by the dynamics of ripples of wind-blown sand, as shown in the next section.

### 3.2 Numerical model of wind-driven clast distributions

To shed light on the physical mechanisms underlying the hyperuniformity, we leverage previous theoretical studies and observations of wind effects on sand surfaces [*Pelletier et al.*, 2009; *Sullivan et al.*, 2022; *Werner*, 1995]. We develop a cellular automata (particle-based) model of clast displacement, sand transport, ripple development, and local erosion-deposition around clasts. The model addresses the displacement of clasts in response to the evolution of the sand surface; causing the local surface slope to change significantly as ripples grow and migrate, and as sand erosion (and deposition) is enhanced upwind (and downwind) of each clast. The numerical model, which has much in common with a previous model of clast displacements on flat sand [*Pelletier et al.*, 2009], elucidates the evolution of clast distributions on rippled sand surfaces (movie S1). It readily yields a diversity of clast distributions very similar to those found on sand surfaces at various sites both in Gale crater and the Gobi Desert, and expected to be common in other stony desert regions.

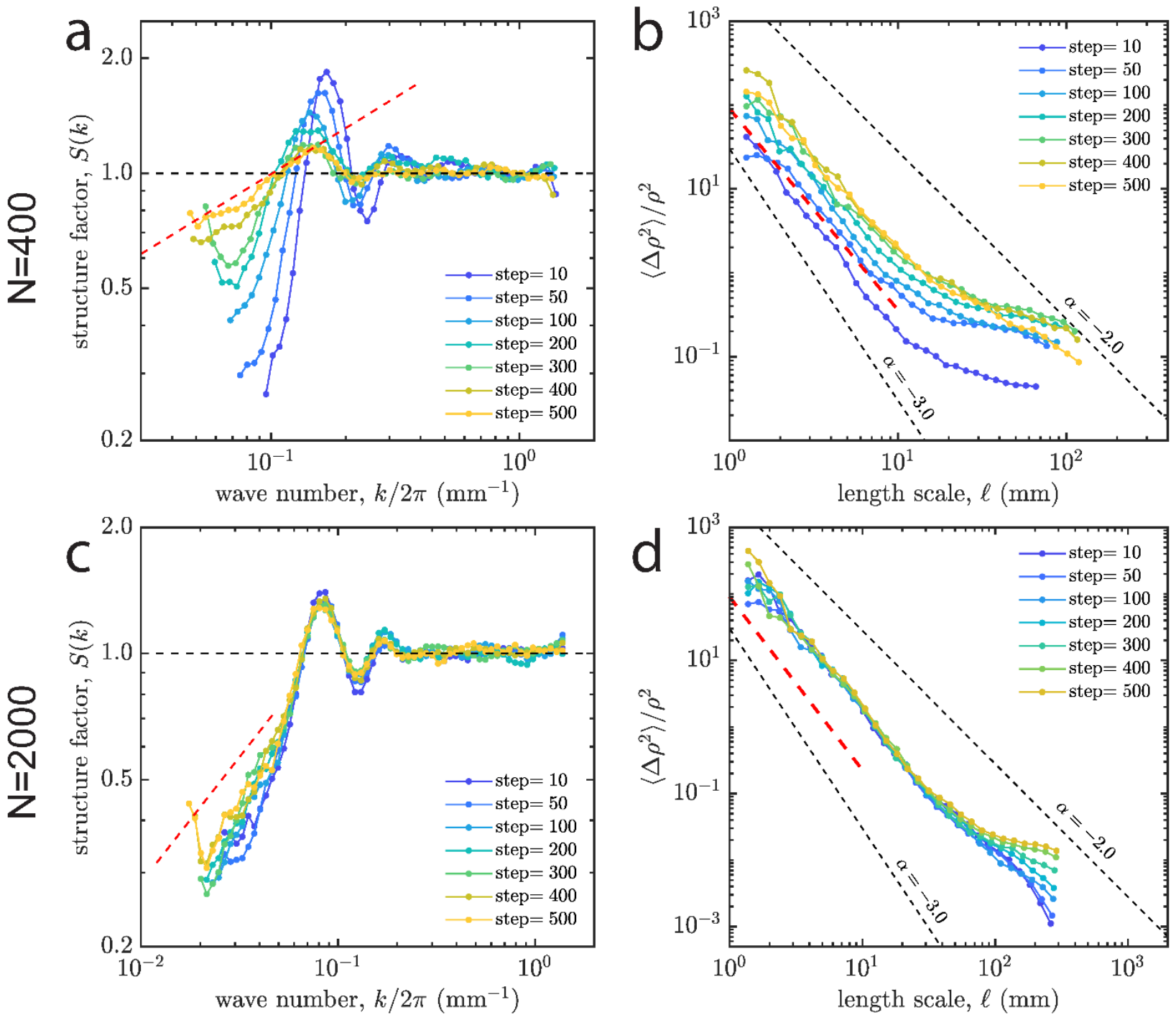

**Fig. 4. Simulated spatial distributions of clasts. Simulations show the time evolution of structure factors and density fluctuations of distributions with different numbers of clasts, $N = 400$ and $2000$.** (**a**, **b**) The exponent of the static structure factor converges to $\gamma = 0.4$ (red dashed line) with increasing time steps in (a), corresponding to an exponent $\alpha$, close to $-2.4$ for the density fluctuations (red dashed line) in (b). (**c, d**) The static structure factor converges to $\gamma = 0.6$ (red dashed line) in (c), which corresponds to $\alpha \sim -2.6$ (red dashed line, d). Simulation results are in good agreement with observed data, all reflecting local hyperuniformity within a range 5 to 10 times the clast radius.

Results from two simulations are presented, each having different numbers of clasts and different areal coverages of clasts initially distributed randomly by the Poisson process: 1) 400 clasts covering ~53% of a square area (Fig. 4a and 4b), and 2) 2000 clasts covering ~36% of a rectangular area (Fig. 4c and 4d and movie S2). For the simulations in Fig. 4, the length and width of the modeled sand surface were set to 1024, a value nondimensionalized by the mean clast diameter (2 mm in Gale, see Fig. S3; 20 mm in the Gobi). The final simulated clast distributions are quantitatively consistent with distributions of actual clasts imaged on Mars and Earth. They all yield a local disordered hyperuniform range (Fig. S8), which can extend to ~10 clast radii from the boundary (indicated by the white circle). Over the course of the simulation, they show similar scaling for both structure factors (Figs. 3a and 4b) and density fluctuations (Figs. 3b and 4b); the exponent $\gamma$ of the structure factor gradually decreases over time (in terms of number of time steps) and then stabilizes near a value of 0.3, indicating long-term behavior (Fig. S9).

The density fluctuations of the imaged (both in Gale and the Gobi) and modeled clasts are consistent with one another; they both yield structure factor exponents that converge to a value around 0.6 (Figs. 3d and 4c), which corresponds to an exponent of about $\alpha$ =-2.6 for the density fluctuations (Figs. 3e and 4d). The spacing between adjacent clasts tends to increase with ripple wavelength in both the field data and simulations, consistent with their shared dependence on the local wind-velocity and saltation lengths (Figs. S10 and S11).

To further explore potential self-organizing aeolian processes, two types of level sand surfaces were considered, one with active ripples (Fig. S12 and movie S1), and the other without [*Pelletier et al.*, 2009]. Models with dynamic ripples lead to disordered hyperuniform clast distributions more readily than those without ripples. The ripples make clasts particularly susceptible to lateral displacements and diffusion as clasts tend to roll stochastically into windward troughs (Figs. S13 and S14). The models show how disordered hyperuniformity develops quickly at local scales and more slowly as the scale increases (Fig. 4 and S8). The progressive self-organization is slow, hence hyperuniformity is only partially achieved at the scale of the entire modeled domain during the simulations. Note that for the hyperuniform state to emerge in our numerical simulations, the clast diameter had to be at least 9 times that of the sand particles, as shown in Fig. 5. Amazingly, this numerical finding, is consistent with the pioneering field observations of dispersions of pebbles on sand by Bagnold [*Bagnold*, 1941]; he wrote that for clasts to disperse evenly on sand "the ratio of the two peak diameters [of clasts and sand] should exceed 10 to 1."

**3.3 A phase diagram for clast distributions and patterns**

Although we have focused on hyperuniform and random clast distributions, other clast arrangements, especially alignments or bands of clasts transverse to the dominant wind direction, are common in nature (Figs. 5g and 5h). Our numerical model offers clues about this diversity of clast distributions that form spontaneously starting from initial random dispersions. Two physical parameters — the saltation distance and the clast-sand diameter ratio — control the density fluctuations exponent of clast distributions and determine the transitions between three distinct types of clast distributions, shown in a novel phase diagram (Fig. 5a): random (blue), hyperuniform (green) and distributions featuring clast alignments (purple).

To explore the implication of the model results (Figs. 5a through 5d), it is useful to recall the scaling exponent ($\gamma$) for structure factors, which determines the dominant type of emerging clast distribution across different wind and surface conditions (see Methods -- *Cellular model of clast displacements on a dynamic sand surface* for details). Within the purple domain, two distinct types of alignment emerge at large saltation distances: when clast-to-sand diameter ratios are also large, they form series of individual parallel alignments or clast lines (Fig. 5, b-I and g), whereas at smaller diameter ratios, they form transverse clusters or clast bands as shown in Fig. 5 b-II and 5 h. The migration of ripples and clasts commonly leads clasts to align along ripple crests (see movies S1 and Fig. S13). The frequency of transverse clast displacements significantly increases when ripples develop (see Fig. S14). In the green domain, disordered hyperuniform distributions emerge; the clast-sand diameter ratio must be at least ~9: 1 (Fig. 5a). We attribute this restriction for hyperuniformity to develop to the interaction of ripple crests and clast movements, whereby ripple dynamics facilitate the lateral migration of clasts.

Fig. 5b to d illustrate simulated clast patterns corresponding to varying values of $\gamma$, and Fig. 5e to h show actual clast patterns recorded in images from Mars, highlighting the considerable similarity between actual and simulated patterns. Our model helps understand the diversity of pebble distributions on aeolian sand surfaces on Mars and Earth, and perhaps on other planetary surfaces.

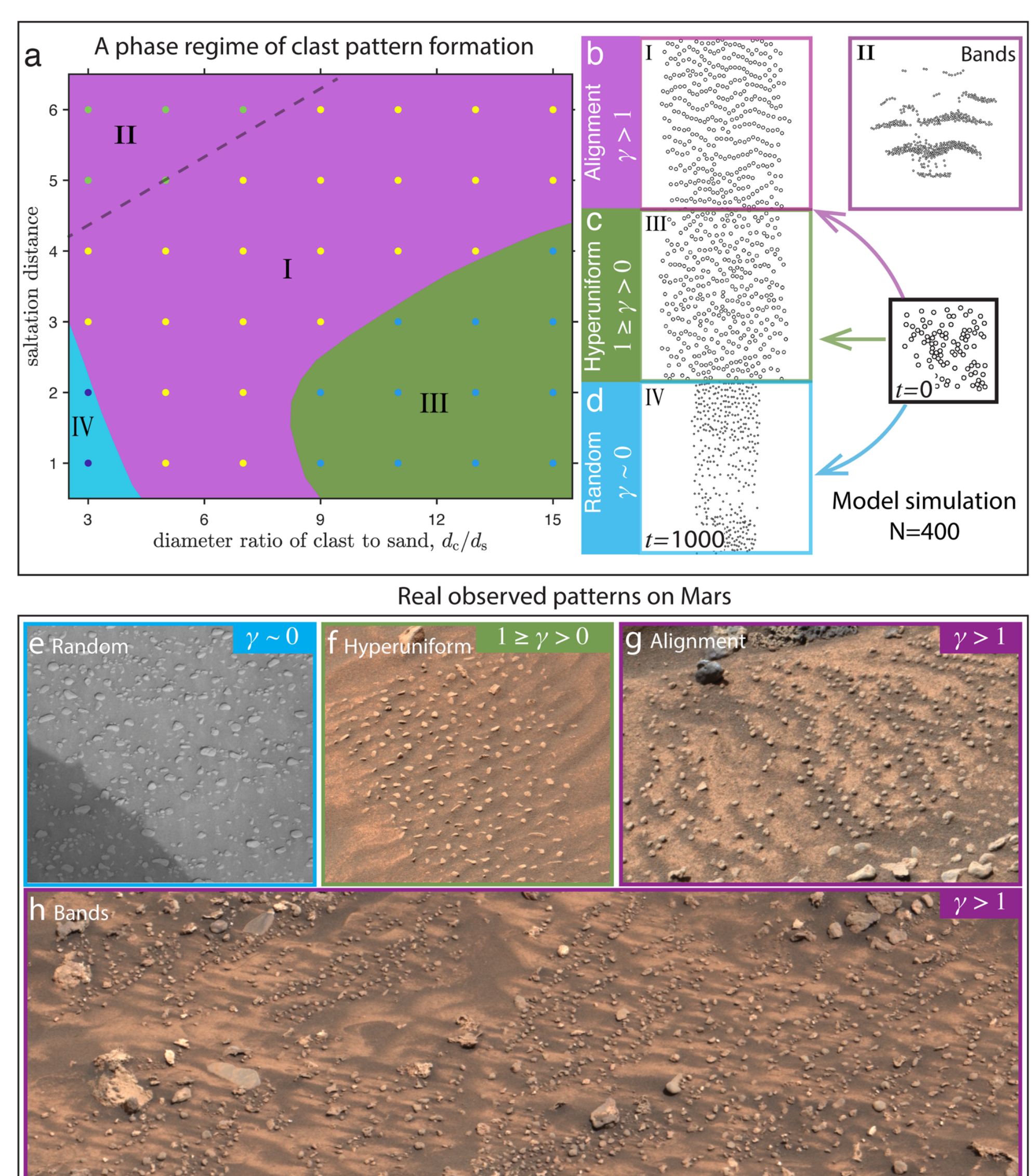

**Fig. 5. Model-predicted regimes of clast distributions governed by the saltation distance and clast-to-sand diameter ratio.** (**a**) Phase diagram for simulated clast distributions with $N = 400$. The density fluctuations exponent $\gamma$ is used to classify the clast distribution regions. The circles represent the simulation parameters of the saltation distance and the diameter ratio. (**b-d**) Examples of spatial distributions starting from the same initial spatial configuration, demonstrating three distinct regimes: alignment and stripes (b), hyperuniform (c), and random distribution (d) at $t = 1000$. (**e-h**) Observed pebble distributions resembling the simulation outcomes: random distributions (e, Sol 2320), hyperuniform distributions (f, Sol 3405), alignments (g, Sol 3737), and bands (transverse clusters) (h, Sol 4229). Image credits: (e to h) NASA/JPL-Caltech/MSSS.

## 4. Discussion

Numerical simulations shed light on how clasts interact with their neighbors through their collective effects on the near-surface wind field and resulting patterns of sand erosion and deposition [*Pelletier et al.*, 2009]. Moving ripples alter clast displacements, locally deviating and accelerating the slow, generally upwind clast migration. This supports the interpretation that Martian ripples, which are known to be active in the current climate, play a larger role in shaping coarse surface materials than previously recognized. The ripples reported herein, from both the Mars images and simulations are centimeter scale, consistent with impact ripples, where grains synchronize with the topography, rather than aerodynamic ripples, which involve a phase shift between turbulent flow and topography [*Yizhaq et al.*, 2024].

Our results suggest that disordered hyperuniformity is a novel stable state for dispersed clasts on sand between random Poisson distributions and alignments (Fig. 5); it requires sustained, moderate wind speeds and suitable grain sizes with clast/sand diameter ratios approaching or exceeding 9, pointing to wind as not only a transport agent, but also a driving force for spatial ordering of clasts larger than can be wind transported on planetary sand surfaces. This organization is important in governing the evolution of surface roughness, influencing boundary-layer wind profiles and potentially stabilizing pebble-rich surfaces against further modification, even in Martian dust cycles. The emergence of hyperuniformity may serve as a natural feedback that reduces local sand transport, thereby limiting further aeolian activity in clast-rich regions. This idea may have practical implications for extensive efforts in western China to mitigate wind erosion in most recent years (Fig. S15).

Our model helps understand the diversity of pebble distributions on aeolian sand surfaces on Mars and Earth, and perhaps on other planetary surfaces. This model and earlier work [*Pelletier et al.*, 2009] represent modest first steps in understanding the complex development of actual aeolian textures and clast dispersions. We have implicitly assumed that the clast distributions we study are currently active and produced by contemporary sand processes and steady wind directions, but we stress that more complex scenarios are to be expected, in general. Clast distributions seen in the rock record, for instance, necessarily involve burial and exhumation and, hence a history involving temporal variations wind strength, orientation, and sediment flux. And yet, so far, we have only considered steady, unidirectional winds, while recent numerical work on aeolian dune forms has highlighted the rich diversity of form resulting from changing wind directions and regimes [*Rubin et al.*, 2024]. Future work promises to expand the diversity of modeled clast distributions, and to improve the interpretation of clast distributions in nature in a range of settings.

More generally, sand transport and related eolian process commonly generate clear spatial variability in the texture of desert surfaces and bedforms, including concentrations of large grains at the crest of large ripples, where they can form a continuous cover with adjacent grains in contact with another; they can also stack to form the crestal region of larger bedforms [*Sullivan et al.*, 2022; *Tholen et al.*, 2022; *Yizhaq et al.*, 2024]. These clast concentrations contrast with the domains of dispersed clasts examined herein in which clasts very seldom contact one another, and that are on sand interspersed in rocky terrain near scattered sources of rock fragments adjacent to nearby bedrock exposures. Our model also indicates that hyperuniform clasts can significantly suppress ripple migration and sand transport. These findings may suggest a potential strategy for using spatially hyperuniform clasts on sand to protect engineering structures, such as highways, from sand transport and burial [*Jerolmack* *and Mohrig*, 2005; *Pellegrino and Prestininzi*, 2007].

Our numerical simulations show that sand ripples, hyperuniform clast distributions, and alignments can all arise spontaneously from steady but stochastic, wind-driven sand grain impacts and subsequent grain motion over uneven sand surfaces. This passive mechanism -- contrasting with active movement in biological systems -- is also supported by independent particle simulations in quite different systems that share similar underlying principles. For instance, in hydrodynamics, long-range interactions are not required for the formation of hyperuniform patterns [*Torquato*, 2016], and the migration of ripples driven by stochastic forces on clasts functions much as in periodically-driven emulsions [*Weijs et al.*, 2015]. Moreover, our simulations help understand the information coded in clast domains. They suggest that the formative winds were perpendicular clast alignments, as well as ripples, and that wind speeds were sufficient to entrain sand, but not too high, otherwise clasts randomize. They also suggest why hyperuniform clast distributions are rare: they require uncommon circumstances such that suitable appropriate wind and ground surface conditions co-occur.

## 6. Summary

We have drawn attention to meter-scale uniform domains of dispersed clasts on sand, a form of self-organization on Mars and Earth that may be widespread, but only rarely recognized as they are considerably more subtle than sand dunes and ripples, and distinct geometric ground patterns [*Hallet et al.*, 2022]. Furthermore, on Mars they are too small to be detected from orbit; they can only be imaged from rovers. Guided by recent advances in fundamental fields --physics, material science and biology-- we utilize concepts and quantitative descriptors (structure factor and pair-correlation function that appear to be novel in aeolian studies) of the spatial distribution of clasts on sand surfaces on Gale Crater, Mars and at analog sites in the Gobi Desert. This approach may also prove useful in studies of diverse other clast distributions on various surfaces on earth and other planets: sediment and regolith surfaces, erosional lags, clast armors and (desert) pavements. Moreover, the descriptors facilitate quantitative comparison between empirical data from various sources, as well as results of numerical models used to probe formative. Lastly, we present our results leveraging rich insights into manifestations of hyperuniformity gained recently in diverse disciplines.

Although our study initially targeted pebble distributions on Mars, it led to unprecedented, quantitative characterization of pebble distributions on Earth. Inspired by many studies that have revealed forms of optimality in diverse other hyperuniform systems, we suggest a hypothesis suitable for future testing: for a given concentration of clasts on sand, the spatial distribution of clasts that armors the sand most effectively is hyper-uniform. This may help understand both the rare occurrence of hyperuniform clast patches, and the transient nature of less uniform patches.

The self-organization of clasts is fundamentally intriguing; it is a visible example of a multitude of inanimate particles collectively acting in an organized manner [*Hazen and Wong*, 2024; *Wong et al.*, 2023]. Moreover, our study methods and findings potentially have implications for significant contemporary societal issues of global significance on Earth, including dust emission into the atmosphere from vast arid areas where clasts and desert pavements can effectively armor the surface and reduce emission [*Meijer et al.*, 2020]. They also bear on the interpretation of dust (loess) accumulation as valuable stratigraphic archives of past changes in environmental factors influencing dust production and emission into planet's atmosphere [*Kim et al.*, 2024; *Z Zhang et al.*, 2022].

## Author contributions

Z.Z. contributed to the analysis, modelling, and performed the simulations. B.H. A.Y., and Q.X.L. conceptualization and data collection. A.A.S. and G.D. contributed to statistical analysis, commented on the paper. Z.Z., B.H. and Q.X.L. contributed to interpretation of the results and writing of the manuscript.

## Competing interests

The authors declare no competing interests.

## Acknowledgements

We are grateful for NASA's support of the Mars Science Laboratory (MSL) mission and for the efforts of the large team of talented and dedicated scientists and engineers who made this work possible. We thank T. Kubacki and his Malin Space Science Systems (MSSS) colleagues for their astute observations and critical role in acquiring the images. MSSS kindly provided the image mosaic used in Figs. 1a and 5 (e to h). We are also thankful for the Clast surveys conducted by the Mars Exploration Rover, Spirit (MER). The authors also appreciate insightful discussions with H. P. Zhang, B. Werner, R. Sletten, R. Sullivan, and M. Malin that benefitted this paper. We thank G. C. Cheng and J. Chen for their assistance in acquiring the drone images and collecting the field data (Tengger Gobi). Funding was provided by the National Science Foundation of China through award No. 32071609 (Q.X.L.), by the NKFIH Grant 149429 and the TKP2021-BME-NVA-470 02 program (G.D. and A.A.S.). B.H. and A.Y. gratefully acknowledge sustained funding through the MSL mission from a NASA grant awarded to MSSS.